\documentstyle[11pt]{article}

\catcode`\@=11
\long\def\@makefntext#1{
\protect\noindent \hbox to 3.2pt {\hskip-.9pt
$^{{\ninerm\@thefnmark}}$\hfil}#1\hfill}                

 \def\@makefnmark{\hbox to 0pt{$^{\@thefnmark}$\hss}}  
 
\def\ps@myheadings{\let\@mkboth\@gobbletwo
\def\@oddhead{\hbox{}
\rightmark\hfil\ninerm\thepage}
\def\@oddfoot{}\def\@evenhead{\ninerm\thepage\hfil
\leftmark\hbox{}}\def\@evenfoot{}
\def\sectionmark##1{}\def\subsectionmark##1{}}
 
\newcounter{sectionc}\newcounter{subsectionc}\newcounter{subsubsectionc}
\renewcommand{\section}[1] {\vspace{0.6cm}\addtocounter{sectionc}{1}
\setcounter{subsectionc}{0}\setcounter{subsubsectionc}{0}\noindent
	{\bf\thesectionc. #1}\par\vspace{0.4cm}}
\renewcommand{\subsection}[1] {\vspace{0.6cm}\addtocounter{subsectionc}{1}
	\setcounter{subsubsectionc}{0}\noindent
	{\it\thesectionc.\thesubsectionc. #1}\par\vspace{0.4cm}}
\renewcommand{\subsubsection}[1] {\vspace{0.6cm}\addtocounter{subsubsectionc}{1}
	\noindent {\rm\thesectionc.\thesubsectionc.\thesubsubsectionc.
	#1}\par\vspace{0.4cm}}

\newcounter{appendixc}
\newcounter{subappendixc}[appendixc]
\newcounter{subsubappendixc}[subappendixc]

\renewcommand{\appendix}[1] {\vspace{0.6cm}
	\refstepcounter{appendixc}
	\setcounter{figure}{0}
	\setcounter{table}{0}
	\setcounter{equation}{0}
	\renewcommand{\thefigure}{\Alph{appendixc}.\arabic{figure}}
	\renewcommand{\thetable}{\Alph{appendixc}.\arabic{table}}
	\renewcommand{\theappendixc}{\Alph{appendixc}}
	\renewcommand{\theequation}{\Alph{appendixc}.\arabic{equation}}
	\noindent{\bf Appendix \theappendixc #1}\par\vspace{0.4cm}}

\renewenvironment{thebibliography}[1]
	{\begin{list}{\arabic{enumi}.}
	{\usecounter{enumi}\setlength{\parsep}{0pt}
\setlength{\leftmargin 1.25cm}{\rightmargin 0pt}
	 \setlength{\itemsep}{0pt} \settowidth
	{\labelwidth}{#1.}\sloppy}}{\end{list}}
 
\newcounter{itemlistc}
\newcounter{romanlistc}
\newcounter{alphlistc}
\newcounter{arabiclistc}

\newcommand{\fcaption}[1]{
	\refstepcounter{figure}
	\setbox\@tempboxa = \hbox{\tenrm Fig.~\thefigure. #1}
	\ifdim \wd\@tempboxa > 6in
	   {\begin{center}
	\parbox{6in}{\tenrm\baselineskip=12pt Fig.~\thefigure. #1}
	    \end{center}}
	\else
	     {\begin{center}
	     {\tenrm Fig.~\thefigure. #1}
	      \end{center}}
	\fi}
 
\newcommand{\tcaption}[1]{
	\refstepcounter{table}
	\setbox\@tempboxa = \hbox{\tenrm Table~\thetable. #1}
	\ifdim \wd\@tempboxa > 6in
	   {\begin{center}
	\parbox{6in}{\tenrm\baselineskip=12pt Table~\thetable. #1}
	    \end{center}}
	\else
	     {\begin{center}
	     {\tenrm Table~\thetable. #1}
	      \end{center}}
	\fi}
 
\def\@citex[#1]#2{\if@filesw\immediate\write\@auxout
	{\string\citation{#2}}\fi
\def\@citea{}\@cite{\@for\@citeb:=#2\do
	{\@citea\def\@citea{,}\@ifundefined
	{b@\@citeb}{{\bf ?}\@warning
	{Citation `\@citeb' on page \thepage \space undefined}}
	{\csname b@\@citeb\endcsname}}}{#1}}
 
\newif\if@cghi
\def\cite{\@cghitrue\@ifnextchar [{\@tempswatrue
	\@citex}{\@tempswafalse\@citex[]}}
\def\citelow{\@cghifalse\@ifnextchar [{\@tempswatrue
	\@citex}{\@tempswafalse\@citex[]}}
\def\@cite#1#2{{$\null^{#1}$\if@tempswa\typeout
	{IJCGA warning: optional citation argument
	ignored: `#2'} \fi}}

\def\fnt#1#2{\footnotetext{\kern-.3em
	{$^{\mbox{\sevenrm #1}}$}{#2}}}
 
 1
 1
 1

\font\tenrm=cmr10

\font\ninerm=cmr9

\begin{document}

\vspace{15 mm}
\begin{center}
\large{{\bf Thermodynamics, topology and dimension of initial
real tunneling manifolds}}
\end{center}

\vspace{5 mm}
\hspace{2 cm}  J.L. ROSALES \footnote{E-mail: rosales@phyq1.physik.uni-freiburg.de}

\hspace{2 cm}
     {\em Fakult\"at f\"ur Physik, Universit\"at Freiburg,}

\hspace{2 cm} {\em Hermann-Herder-Strasse 3, D-79104 Freiburg, Germany}

\vspace{10 mm}

\begin{quote}
\begin{center}
				Abstract
\end{center}
	
	Based on the Non-Boundary proposal in quantum cosmology,
	we  develop the argument that initial real
	tunneling in quantum gravity might be 
	contemplated as a thermodynamical analogous
	to a black hole condensate in equilibrium with Hawking's 
	radiation in a box. The total entropy is always maximized 
	in the Lorentzian
	sector of the theory, and, in this
	sense, tunneling is predicted. 
	The maximum relative increase of the entropy 
	is achieved if the Euclidean geometry has topology 
	$\cal{S}$$^{2}$x$\cal{M'}$$^{2}$  
	($\cal{M'}$$^{2}$ so far an arbitrary bidimensional manifold)
	and the total spacetime dimension is four.
	We propose that a 
	{\it tunneling from nothing} configuration
	in quantum cosmology  be contemplated
	as an initial condition for the Universe. 

\end{quote}
\vspace{2 mm}

\section{Introduction.}

Real tunneling manifolds \cite{kn:Gibbons1} are a very restrictive set of
solutions of Euclidean Einstein's equations with positive definite
metrics, compact topologies and finite actions 
being  the "complexified" 
spacetimes of some suitable real Lorentzian signature metrics solving 
the field equations with a new time coordinate  $t=i\tau$ ($\tau$ standing for
the $x^0$ coordinate in the Euclidean geometry). 
These have been described  as gravitational instantons \cite{kn:Eguchi}.
On the other hand,
a general positive metric will not have a section on which the metric is real
and Lorentzian. The importance of these metrics comes from the fact that 
they  must dominate in the path integral defining 
the wave function of a closed Universe \cite{kn:Hawking1}.
Moreover, a real tunneling solution describes transitions from a purely 
"Euclidean" metric to a purely Lorentzian (i.e., a 
"tunneling from nothing" configuration  \cite{kn:Vilenkin1}).
In this scenario, the creation of the Universe "ex nihilo" is due to the 
presence of an effective cosmological constant and the result may also be 
considered as a sort of classical change of signature  or a bounce in spacetime 
(i.e., the transition between two solutions with the same boundary conditions 
having different actions). In quantum cosmology, a tunneling solution of the 
Wheeler-DeWitt equation in minisuperspace with a positive cosmological constant 
would represent the quantum rate of production corresponding to a spacetime of
the type of a round Euclidean sphere joined on an equator to the Lorentzian
space at its radius of maximum contraction \cite{kn:Gibbons1}.
 
Based on the Non-Boundary proposal\cite{kn:Hartle} 
we can also visualize  gravitational instantons as having thermodynamical
properties rather like a black hole. Though, it is not obvious from the behaviour of the wave
function in the saddle points approximation, 
$\Psi\sim \exp[-I]$ (here, $I$ stands for 
Einstein's Euclidean  action depending on the topology). 
On the other hand,  semiclassically, one can asign a probability measure to
the compact instanton simply as the square of the
wave amplitude for the Universe, i.e.,  $P=|\Psi|^2\sim \exp\{-2 Re[I]\}$.
It leads to  computing  the logarithm of the number of 
available (microcanonical) states as in the case 
of a thermal system with an entropy rather like
$S=\log (P)\sim -2 Re[I]$, 
which is finite and non vanishing due to the fact  that the topology of 
the instanton is  not trivial, a feature which also holds
in the black hole case\cite{kn:Hawking2} 
(a quite direct and interesting example  of this is 
observed in four dimensions
when there exists a positive cosmological 
constant, $\lambda$, so that $I\sim -\kappa\chi/(16 \pi G\lambda) $
--$\chi$ stands
for the Euler-Poincar\'e charactereistic of the instanton 
and $\kappa\leq 12 \pi^2$ is some numerical constant \cite{kn:Hawking3}). 

For a compact manifold there exists obviously no real time, 
so there should not either be
necessarily  an ordinary space. For instance, we might expect 
the available dimensions of the global manifold to be an arbitrary, yet
undetermined, number $n$. 
Moreover, that part of the compact instanton describing an effective 
change of signature having a  Lorentzian sector
will probably consist on a $\cal{S}$$^{d}$ 
($d$, so far, also some free parameter). This is because we expect 
to extend the analogy with black holes to the existence of 
thermal Green functions on the complexified spacetime
having some  real period  equating the inverse of the physical
temperature. It  imposes a periodic topology $\cal{S}$$^{d}$. 
If the geometry is connected the instanton would consist on a  generic topological product,
$\cal{M}$$^{n}\approx$$\cal{S}$$^{d}$x$\cal{M'}$$^{n-d}$. 
Nonetheless, as we will
see, thermodynamical consistence will strongly restrict 
the topology of the instanton.  

For $n$-dimensional  gravitational instantons (when $n>3$)
as well as for black hole thermodynamics one finds
a negative specific heat. This is in conflict with the positivity
of $(\delta \hat{U})^{2}$ for the fluctuations of the thermal 
energy. It means that both systems are thermally unstable. As such this is
not surprising since instability is typical for gravitational 
fenomena\cite{kn:Zeh},
a feature which in turns seems to appear even in Newtonian theory
as showed by Jeans\cite{kn:Jeans}. 
Moreover, one can use the the microcanonical ensemble 
to obtain an equilibrium configuration. This was done previously by
Gibbons and Perry \cite{kn:Gibbons2} who
considered a black hole immersed in a bath of radiation with fixed
volume: they obtained that , at a sufficiently high energy density, a black
hole nucleates from pure radiation in a way analogously to a liquid
drop can condense out of saturated vapour. From  the similarities
existing between black holes and  gravitational instantons, 
the previous heuristic
picture  may  also be seen as useful in order to understand the physical
phenomenon which underlies in a generic {\it tunneling configuration}
even in  the cosmological case (so  that, the phenomenon of  {\it creation of
the Universe from nothing}  be compared 
to that of a condensed liquid drop  from saturated vapour).

Quantum gravity is the available technology of the fabric of spacetime
and, indeed, as every technology does, it is constrained by thermodynamics;
to us, it remains to obtain the answer to the question: 
{\it what is the topology of the Euclidean instanton from which a stable 
thermal Universe may have been nucleated using this technology?} 
This problem is admitedly elusive, a fully satisfactory solution
would require a more convincing theory of quatum gravity, here,
however, we may try to deal with some simplified arguments, 
in the spirit of  the semiclassical  approximations, 
so that the question be  affordable.

The paper is organized as follows, in section 2 we review the 
thermodynamical  properties of those gravitational 
instantons corresponding to "tunneling
from nothing configurations". In section 3 we develop the analogy between
thermal equilibrium of a black hole made of pure radiation inside a box
and the nucleation of a initial tunneling manifold. Section 4 is devoted
to obtaning the "entropically favoured states", that is, the topology
and dimension corresponding to the maximization of a  relative 
entropy function, already defined in section 3, 
upon considering the Boltzmann thermal 
equilibrium of the nucleating manifold with the radiation in
an arbitrary space-like cavity. Our final conclusions are written
in last section.

\section{Topology and thermodynamics}

A real tunneling manifold $\cal{M}$$^{n}$ might be also contemplated as a gravitational instanton
in an arbitrary dimensional "complexified" spacetime \cite{kn:Embacher}. In this
case, it is described by means of its classical action and it has a non zero entropy
\begin{equation}
S=-2I(\mbox{$\cal{M}$}^{n}/2) \mbox{,}
\end{equation}
where $I(\mbox{$\cal{M}$}^{n}/2)$ is the classical Euclidean action computed for half of the
compact manifold joined to the Lorentzian geometry at the spacelike hypersurface
of vanishing extrinsic curvature.

The corresponding temperature of the instanton is given by the usual 
thermodynamical formula
\begin{equation}
T^{-1}=\frac{\partial S}{\partial U} \mbox{,}
\end{equation}
$U$ standing for the thermal energy of the instanton. On the other hand, the
temperature is the inverse of the period of the complex time coordinate of a 
d-dimensional sphere of radius $r_d$ immersed into the Euclidean n-dimensional
geometry  $\cal{M}$$^{n}$ in the form of an arbitrary topological product 
$\cal{M}$$^{n}\approx$$\cal{S}$$^{d}[r_d]$x$\cal{M'}$$^{n-d}$, i.e.,
\begin{equation}
\beta=T^{-1}=2\pi r_d \mbox{.}
\end{equation}

In order to compute the action we require the expression for 
the Ricci tensor  of $\cal{S}$$^{d}[r_d]$, it is given by,
\begin{equation}
R_{ab}[\mbox{$\cal{S}$$^{d}$}]=\frac{(d-1)}{r_d^2}g_{ab}[\mbox{$\cal{S}$$^{d}$}] \mbox{,}
\end{equation}
and,
\begin{equation}
R[\mbox{$\cal{S}$$^{d}$}]=\frac{(d-1)d}{r_d^2} \mbox{.}
\end{equation}
On the other hand, Einstein's equations are given in terms of an effective
cosmological constant
\begin{equation}
R_{\mu\nu}[\mbox{$\cal{M}$}^{n}]=\frac{2\lambda}{n-2}g_{\mu\nu}[\mbox{$\cal{M}$}^{n}] \mbox{.}
\end{equation}

Equations (4)-(6) directly impose
\begin{equation}
r_d^2=\frac{(n-2)(d-1)}{2\lambda} \mbox{,}
\end{equation}
that is,
\begin{equation}
\beta=2\pi[\frac{(n-2)(d-1)}{2}]^{1/2}\lambda^{-1/2}
\end{equation}

As a matter of fact, we can not obtain the general expression for the
entropy corresponding to the nucleating manifold (i.e.,the d-sphere, which 
undertakes an effective change of signature in the $x^0$ coordinate when 
analytically continuing the Euclidean metric into the Lorentzian sector). 
This is because, the Euclidean action  also depends 
on the value of Newton's constant in n-dimensions 
$G_n=(k_n/m_Pl)^{n-2}$, and the $k_n$ are arbitrary so far but in the standard 
four dimensional case ($k_4=1$),
\begin{eqnarray*}
S=-2I(\mbox{$\cal{M}$}^{n}/2)=\frac{2}{16\pi G_n}\int_{\mbox{$\cal{S}$$^{d}/2$}}\{g[S^d]\}^{1/2}d^{d}x\int_{\mbox{$\cal{M'}$}}(R-2\lambda)\{g[M']\}^{1/2}d^{n-d}x= 
\end{eqnarray*}
\begin{equation}
=\frac{2}{16\pi G_n}\frac{4}{n-2}\lambda \frac{1}{2}\mbox{$\cal{V}$}_{S^d} \mbox{$\cal{V}$}_{M'}=\zeta(n,d) \frac{\lambda^{-(n-2)/2}}{G_n} \mbox{.}
\end{equation}
We  used the fact that the compactified internal space $\cal{M'}$$^{n-d}$ has
a  volume $\cal{V}$$_{M'}\sim\lambda^{-(n-d)/2}$; it will remain compact 
after the effective change of signature.
The function $\zeta(n,d)$ depends on the topology of the global manifold. 
In the special case that $d=n$, we get 
\begin{equation}
\zeta(n,n)=\frac{v_{n-2}}{4}(\frac{(n-1)(n-2)}{2})^{(n-2)/2} \mbox{,}
\end{equation}
where,
\begin{equation}
v_{n-2}\equiv \frac{2\pi^{(n-1)/2}}{\Gamma((n-1)/2)} \mbox{,}
\end{equation}
hencefore we obtain the result \cite{kn:Gasperini}
\begin{equation}
S=\frac{A}{4 G_n} \mbox{,} 
\end{equation}
$A$ standing for the surface area of the horizon at which
the effective change of signature takes place. It is exactly the same 
expression than the one obtained for black holes. 
It suggests a  deep conection between compact instanton 
thermodynamics and that from black holes,
though, the physical analogy is problematic 
since, in the black hole  case, the area law corresponds to the presence of
an horizon in a non compact Lorentzian spacetime. 

On the other hand, from Eqs. (8)-(9) we can write for the entropy 
\begin{equation}
S=\gamma(n,d)\frac{\beta^{n-2}}{G_n} \mbox{,}
\end{equation}
for $\gamma(n,d)$ a known function of the parameters of the topology of
the compact manifold. Now the thermal energy is simply, 
\begin{equation}
U=F+TS=\int \beta^{-2}S d\beta +\beta^{-1}S =\gamma(n,d)\frac{\beta^{n-3}}{G_n}\frac{n-2}{n-3} \mbox{,}
\end{equation}
and, 
\begin{equation}
S=U\beta\frac{n-3}{n-2}=2\pi r_d U \frac{n-3}{n-2}< 2\pi r_d U \mbox{;}
\end{equation}
which is Bekenstein's bound \cite{kn:Bekenstein}. Let us write the above
expressions in the form
\begin{equation}
S=C(d,n,k_n)[\frac{U}{m_{Pl}}]^{(n-2)/(n-3)}.
\end{equation}
Here, $C(d,n,k_n)$ is  a function of the total dimension and the topology 
which also depends on the physically undertermined quantities $k_n$;
these values should only be restricted by the fact that, 
recalling a classical 
analogy with statistical mechanics,
we expect $S$ to
increase with the total dimension $n$ in a way that should
mimic a comparative increase of the volume of phase space with 
dimensionality.

Heat capacity can not be defined positive for a gravitational instanton, 
therefore,
thermal energy fluctuations become imaginary, 
a signal that the system
is thermally unstable; this is also the case for a generic black hole.
In the later case, however, 
one can consider the  canonical ensemble for a 
hole inside a cavity full of radiation in
Boltzmann equilibrium with it \cite{kn:York}; it leads to a 
thermodynamically stable description 
making possible to understand the nucleation of black holes  
from hot flat space\cite{kn:Perry}.  The latter suggests to extend
the themodynamical analogy of compact instantons  with black holes 
to the Lorentzian sector of the tunneling
manifold where there would also exist radiation and ($d-1$ dimensional) 
volume terms. 

More simply, let us study the microcanonical 
ensemble concerning the maximization of the 
total entropy of the system  with radiation in a {\it fixed} volume at a 
given {\it constant} total energy.

 \section{Thermal equilibrium with radiation}

Correspondingly to the non-vanishing temperature,
we should consider the black body radiation inside an arbitrary 
$d-1$ dimensional cavity of the nucleated  space  having a volume
$V_{d-1}$. 
Boltzmann thermal 
equilibrium is reached for a given volume and fixed total energy configuration
that maximizes the total entropy of the system including the radiation terms,
\begin{equation}
S_T=C(U/m_{Pl})^{(n-2)/(n-3)}+(1+\frac{1}{d-1})T^{d-1}V_{d-1}\sigma(d)
\end{equation}
\begin{equation}
E=U+\sigma(d) V_{d-1} T^d
\end{equation}
Here, $\sigma(d)$ is Stefan's constant in $d$ dimensions and we assume the
standard expressions for a black body gas in a $d-1$ dimensional cavity. 
This consideration is in complete analogy with that from the equilibrium of 
a black hole with Hawking's radiation 
in a box discussed earlier by Gibbons and Perry\cite{kn:Gibbons2}.

Now, upon eliminating $T=[(E-U)/(\sigma_d V_{d-1})]^{1/d}$ from the above equations 
and defining
\begin{eqnarray}
A(E,V_{d-1})=\frac{d}{d-1}(\sigma_d V_{d-1})^{(d-1)/d}E^{1/d}\mbox{;} \\
g(E)=C(\frac{E}{m_{Pl}})^{(n-2)/(n-3)} \mbox{;} \\
\Omega=S_T/g(E) \mbox{;} \\ \omega =A(E,V_{d-1})/g(E) \mbox{;} \\ u=U/E \mbox{,} 
\end{eqnarray}
we can express the total entropy as a function of the previous 
adimensional variables,
\begin{equation}
\Omega(u,\omega)=u^{(n-2)/(n-3)}+\omega (1-u)^{(d-1)/d} \mbox{,}
\end{equation}
$\Omega$ represents the relative total entropy weighted by the entropy of an
instanton of total thermal energy. 
It  does {\it not} depend on the exact topology of the compactified
space $\cal{M'}$$^{n-d}$ neither on the value of the unknown constants $k_n$.
On the other hand, we propose that the energy variable 
$ 0\leq u \leq 1$ had the the following heuristic meaning: 
a configuration such that $u<1$ would represent a tunneling configuration
with some amount of radiation in thermal equilibrium (at some temperature)
with the gravitational field (i.e., the horizon would absorb exactly the same 
energy than it emits).It is in full analogy
with black hole nucleation from hot flat space \cite{kn:Gibbons2} 
\cite{kn:Perry}.
The radiation would have, therefore, 
an energy which is a portion  of the total one, 
$E_{\mbox{{\it rad}}} = E(1-u)$. 
Of course, the solution to $\Omega(u,\omega)=1$ 
is also $u=1$ for all values of $\omega$ (the instanton configuration).

Thus, in the above scenario,  we are led
to a picture which is the cosmological analogous to a black hole 
condensate  in thermodynamical equilibrium with 
pure radiation \cite{kn:Gibbons2},
here, however, the existence of radiation
is a {\it consequence} of the {\it condensation} of a tunneling configuration 
in its Lorentzian sector (since, previously to the {\it creation of the Universe from nothing} it is 
impossible to define radiation in a space-like cavity).
This is conversely to the black hole case in which radiation is previous to
the existence of the hole. Thus $\Omega(1,\omega)=1$ would be the total relative
entropy in the special case that there were no volume terms. 
{\it Nucleation} is entropically 
favoured if and only if there exist a value $u'<1$ such that $\Omega(u',w)>1$.
Therefore, {\it tunneling} would be allowed if the total entropy 
had increased in the Lorentzian sector 
with respect to the entropy defined by means of the topological properties
of the conpact instanton.
In the black hole case, nucleation will imply, in general,  
a spontaneous increase of the free energy $F$ (since
hot flat space in four dimensions have negative $F$ whereas 
black hole  Helmholz's free energy is necessarily positive). 
That difficulty,  however, does not apply to the cosmological case since, 
as we  just mentioned, the standard picture evolution is formally 
reversed. 

In order to find the value  $u'$ that maximizes the relative entropy in 
Eq. (24), we should obtain  $\omega=\omega'$  satisfying the constraint
\begin{equation}
\frac{\partial \Omega}{\partial u}|_{u'}=0
\end{equation}
or,
\begin{equation}
w'=\{\frac{(n-2)}{(d-1)(n-3)}+1+\frac{1}{n-3}\}(1-u')^{1/d}u'^{1/(n-3)}
\end{equation}

On the other hand, the global maximum of $\Omega$ is reached at $u=0$,
unless there were a different value $\Omega(u')$ such that 
\begin{equation}
\Omega(0)=\omega'=u'^{(n-2)/(n-3)}+\omega' (1-u')^{(d-1)/d}=\Omega(u') \mbox{,}
\end{equation}
solving the above equation for $\omega'$ we obtain,
\begin{equation}
\omega'=\frac{u'^{(n-2)/(n-3)}}{1-(1-u')^{(d-1)/d}} \mbox{,}
\end{equation}
now, from Eq. (26) and Eq. (28), we directly get
\begin{equation}
1-u'=[\frac{d(u'-n+2)+u'(n-3)}{d(2-n)}]^d \mbox{,}
\end{equation}
or, upong doing the  definitions 
\begin{eqnarray}
a(n,d)=\frac{d-3+n}{n-2}  \mbox{;} \\
x=u'a(n,d)
\end{eqnarray}
and substituting them in Eq. (29) we finally get:
\begin{equation}
1-\frac{x}{a(n,d)}=[1-\frac{x}{d}]^d \mbox{.}
\end{equation}
The above equation has solutions $x$ different from zero if and only
if $a>1$, that is $d>1$, this is also physically consistent since  
$\cal{S}$$^{1}$ could not represent a tunneling manifold.
For $n=d=4$, we obtain, $a[\mbox{$\cal{S}$$^{4}$}]=5/2$, then, 
$u'=x/a[\mbox{$\cal{S}$$^{4}$}]\approx 0.97702$ is the fraction
of the energy corresponding to the nucleation 
of deSitter spacetime from nothing. Notice that it coincides with
Gibbons and Perry heuristic estimates in the case of a neutral
black hole condensate from pure radiation\cite{kn:Gibbons2}.
The latter amounts to the well known analogy between deSitter's
spacetime and Schwarzschild's black hole thermodynamics.

Yet, since $\Omega(u')$ should  also be the  local maximum,
it satisfies,
\begin{equation}
\frac{\partial^2\Omega}{\partial u^2}|_{\omega',u'}< 0 \mbox{.}
\end{equation}
This directly requires -using Eq.(26)
\begin{equation}
\frac{d}{d+n-3}< u'<1 \mbox{,}
\end{equation}
that is $n >3$.  Therefore, {\it thermodynamics  requires that the minimum number
of total dimensions for a tunneling manifold be four!}

Moreover, it is a simple task to proove that the equilibrium
temperature coincides with the instanton temperature. This can be easily done,
for instance, for the special case that $n=d$ (see Apendix A). 

\section{Topology and dimension}

Since $u=U/E \sim r_d^{n-3}\sim \lambda^{-(n-3)/2}$, a finite value of
$u=u'$ which maximizes the total entropy also represents a finite
value of the effective cosmological constant whose
dynamical meaning is that of a large potential for the
scalar matter field (inflaton). Its value would remain approximately
constant  if we could neglect  back reaction of radiation 
on the gravitational field. In this case  $E_{rad}\ll E$ and $u'\sim 1$. 
This fact could be considered as 
a physical requirement of consistence in the 
quantum cosmological tunneling scenario. Moreover, more straightforwardly,
a solution such that $u'\approx 0$ amounts to $\lambda/m_{Pl}^2 \gg 1$ and
this  clearly breaks down the semiclassical approximations on which the thermal
description was based.

On the other hand, the solutions $u'=ax$ of Eq. (32) 
have the following behaviour for
large $n$ and finite $d$,
\begin{equation}
u'_{n\gg 1, d\sim O(1)}\sim \frac{2d}{n}[1-\frac{4(d-2)}{3n}+ O(n^{-2})]\rightarrow 0 \mbox{.}
\end{equation}
Which seems to be in contradiction with the thermal description after
the previous heuristic reasoning. It means that, at least in
the semiclassical domain of energies, we must consider 
$d\sim O(n)$, in particular, $d=n(1-\alpha)$, for  some
$\alpha\leq \alpha_{max}< 1$ and $n\leq d/(1-\alpha_{max})$.

Now, in order to find the solution of Eq. (32), we need to know the range
of $a(n,d)$ for the allowed tunneling configurations. First, since $n\geq 4$,
we have from Eq. (30), and that $d\equiv n(1-\alpha)$,
\begin{equation}
\frac{1}{n}=\frac{1}{2}[1-\frac{1-2\alpha}{2a-3}]\leq \frac{1}{4} \mbox{,}
\end{equation}
that is,
\begin{equation}
1< a\leq \frac{5}{2}-2\alpha =\frac{1}{2}+\frac{2d(n)}{n} \mbox{.}
\end{equation}
The topology of $\cal{M}$$^{n}$ 
is encoded within the variable $a$, for instance, the 
case $a=3/2$ may correspond to the
topological product of identical spheres $\cal{S}$$^{d}$x$\cal{S}$$^d$ 
whereas $a=5/2$ denotes the $4$-sphere $\cal{S}$$^{4}$. We can assume that 
$a$ takes all possible rational values in its interval and so, it is a 
dense variable. Yet, $a>1$ and  the inequality above also
suggests that $d$ must be some suitable function of $n$ (otherwise 
$\lim_{n\gg 1} a\leq 1/2$, contrary to the assumption
that there exist non zero solutions of Eq. (32)).

Since we would like to find
the entropically favoured topologies, we should  evaluate the
absolute maximum of
$\Omega[u']=\omega'$ at the
solutions of Eq. (32),i.e., $u'=x(a,d)/a$. It is easy to show (see Apendix B) 
that, in the semiclassical regime, i.e., 
when  $u'\sim 1$, $\omega'$ in Eq. (28) is approximated by,
\begin{equation}
\omega'\sim \exp\{\frac{1}{da^3}\}+O[(1-u')^2] \mbox{,}
\end{equation}

Moreover, the simpler approximation to the solution of Eq. (32) is to
take $x=a u'\sim a$ independent of $d$; a much better approximation is
(see also Apendix B)
\begin{equation}
x\sim a \exp\{-\frac{1}{a^3}[1-\frac{a}{d}+O[(1/d)^2]]\} \mbox{,}
\end{equation}

Since $\cal{S}$$^{1}$ is not a tunneling manifold, we have $d\geq 2$ and
recalling Eq. (30),
\begin{equation}
d=n(a-1)-2a+3\geq 2
\end{equation}
namely,
\begin{equation}
a\geq \frac{n-1}{n-2} \mbox{,}
\end{equation}
but, using Eq. (37), also recalling that thermal (semiclassical) 
description  requires $d=d(n)\sim O(n)$, we get
\begin{equation}
\frac{n-1}{n-2}\leq \frac{1}{2}+\frac{2d(n)}{n} \mbox{,}
\end{equation}
obtaining, 
\begin{equation}
d(n)\geq \frac{n^2}{4(n-2)} \mbox{.}
\end{equation}

Eq. (43) could be  considered as an aditional physical constraint for
the effective $d$-dimensional signature change spacetime geometry. Thus, 
for instance, the maximum total dimension of the Euclidean manifold having a Lorentzian
$4$-dimensional sector could be estimated to be $n_{max}=13$. Yet, this is 
in accordance with Embacher's related previous calculations 
for the available Euclidean geometries in quantum gravity, using
the procedure of minimizing Einstein's action for the Euclidean manifold
corresponding to different dimensions having various 
topologies \cite{kn:Embacher} .

Feeding $d=n(1-\alpha)$ in Eq. (43) we obtain 
\begin{equation}
\alpha_{max}< 3/4 \mbox{.}
\end{equation}

In order to find $\alpha_{max}$, we may
consider the following physical conderation:
if a generic  $n-d$ dimensionsional 
space remains compactified during
cosmological evolution, there would exist, associated to it,
an inverse temperature whose maximum possible (non equilibrium) 
value is  
in terms of  the radius of a maximal $n-d$-sphere, $\cal{S}$$^{n-d}$
\begin{equation}
\beta_{c}=2\pi r_{n-d} \mbox{,}
\end{equation}
where $r_{n-d}$ is,
\begin{equation}
r_{n-d}^2=\frac{(n-2)(n-d-1)}{2\lambda} \mbox{,}
\end{equation}
The previous simplification 
does not seem to be relevant to evaluate $\Omega$ 
for recall that $\Omega$ does not depend on the exact topology 
of $\cal{M'}$$^{n-d}$.

Yet, the cosmological horizon holds an inverse temperature
given by Eq. (3) and Eq. (7) and, in order to prevent it
from evaporating out (so that the horizon 
radiates less energy than it absorbs from extra dimensions)
it is  required that the compactified space 
had a highest possible associated inverse temperature satisfying,
\begin{equation}
\beta_{c}\leq \beta_{Horizon}=\beta(r_d) \mbox{,}
\end{equation}
the latter directly amounts to
$n-d\leq d$, i.e., $\alpha_{max}=1/2$. It is compatible with the constraint 
in Eq. (44).
Moreover, the compactified
space will remain "small" during cosmological evolution until
it eventually reaches some Planckian size whereas cosmological horizon
inflates.

We have finally reached the state on which we would be able to give
an answer to the question which is the aim of this paper: in the 
semiclassical inflationary regime corresponding to the Lorentzian
sector of a generic initial real tunneling manifold, the only available 
Euclidean topologies compatible with a stable thermal regime 
are  $\cal{M}$$^{n}\approx$$\cal{S}$$^{d}$x$\cal{M'}$$^{n-d}$, where 
$dim[$$\cal{M}$ $]$$\leq 2d$, and $d$ is such that there exist an 
absolute maximum for $\omega'$ written in Eq (38). The exact topology of  
$\cal{M'}$$^{n-d}$ remains in principle arbitrary.

\vspace{2 cm}

\section{Conclusions}

A maximal increase of the relative entropy
is obtained when  the variable $a$ reaches its minimum value, i.e., for $a=3/2$. 
The latter encodes that the Euclidean topology of the tunneling manifold should 
be $\cal{S}$$^{d}$x$\cal{M'}$$^{d}$. 
The absolute maximum is now for the minimal
available dimension $d=2$ of the effective signature change geometry 
so that the topology of the compact instanton
is $\cal{S}$$^{2}$x$\cal{M'}$$^{2}$. The total dimension of 
spacetime should be four, at least for the range of energies 
compatible with the semiclassical thermal approximations. 
The previous beautiful, although admitedly heuristic,
result indicates that a typical physical "a priori" such as the dimension
of spacetime should  not be assumed but derived from more developed theories
of gravity such as string theory but that, since Einstein's gravity should be recovered
at low energies, the compactification to four dimensional spacetime
have to take place by consistence of the semiclassical theory as 
we have stablished here.

On the other hand, Eq. (38) states that a relative increase 
of the total entropy is always achieved in the Lorentzian sector of
a tunneling manifold. This suggest that a "tunneling
from nothing" configuration should be considered 
as an initial condition for the Universe 
in the sense of thermodynamics, i.e., that there should exist an
{\it arrow of time} in the direction of increased entropy.
It also means that tunneling 
solutions can be thought as initial states in cosmology. 
Conversely, a "collapse into
nothing" configuration seems to be strongly 
thermodynamically prohibited by these (semiclassical) estimates. 

The latter might  be  seen as an argument against
the elaborated comments of Kiefer and Zeh \cite{kn:KieferZeh}
about the non existence of a generic arrow of time in cosmology.
This is not the case.
The author agrees with the former
in their line of thought concerning the partial effectiveness 
of cosmological decoherence in order to define 
an arrow of time. The latter is true in particular for the
solutions of Wheeler-DeWitt equation which could not (in principle) 
be used to describe a formal evolution in time,
but, on the other hand, the existence of a non trivial entropy
seems to be strongly related on  peaking up some suitable 
boundary condition for the wave function of the Universe (in this case
the Hartle-Hawking state). 
That is why, indirectly, a thermal picture, 
being not independent on the selection of the boundary condititions, arises.
In order to derive consistence one can device simple
{\it Gedankenexperimente} as the one we have suggested here.
In general we should  never ignore the very existence of 
some, perhaps, absolute (i.e., topological) entropy  in the discussions, 
a fact which should be in the core of quantum 
gravity \cite{kn:Penrose}.

\section{Acknowledgments}

The author wishes to thank the Department of Physics of the University
of Freiburg for his hospitality. This work is supported by a postdoctoral
grant from the Spanish Ministry of Education and Culture and
the research project C.I.C. y T., PB 94-0194.

\vspace{1 cm}
\noindent {\bf Apendix A}

\vspace{3 mm}
Let us proove that the equilibrium temperature is exactly that
of the  instanton in the special simpler case that the topology of the
compact manifold is $\cal{S}$$^{n}$. In this case we have to maximize
$S_T$, at fixed total energy (we set $E=1$ for convenience) and
fixed volume $V_{n-1}$.
\begin{eqnarray*}
S_T=\frac{1}{4}v_{n-2}r_n^{n-2}+\frac{d}{d-1}\sigma_n V_{n-1}T^{n-1}    \\
E=1=\frac{r_n^{n-3} (n-2)v_{n-2}}{8 \pi (n-3)}+\sigma_n V_{n-1}T^n \mbox{;} 
\end{eqnarray*}
\hspace{12 cm } {\bf A.} 1

\vspace{1 mm}
We now define the quantities 
$A=n/(n-1)(\sigma_n V_{n-1})^{1/d}$, $C=v_{n-2}/4$, 
$B=v_{n-2}(n-2)/(8\pi(n-3))$ and $u=B r_n^{n-3}$; so that 
the entropy of a generic tunneling configuration be written as
\begin{eqnarray*}
S(1)=\frac{C}{B^{(n-2)/(n-3)}} \mbox{,} 
\end{eqnarray*}
\begin{eqnarray*}
S_T=S(1)u^{(n-2)/(n-3)}+A(1-u)^{(n-1)/n} \mbox{.}
\end{eqnarray*}
\hspace{12 cm } {\bf A.} 2

\vspace{1 mm}
Yet, the equilibrium temperature is from {\bf A.}1,
\begin{eqnarray*}
T_{0}=[\frac{1-u}{\sigma_n V_{n-1}}]^{1/d}=\frac{n(1-u)^{1/n}}{(n-1)A} \mbox{,}
\end{eqnarray*}
\hspace{12 cm } {\bf A.} 3

\vspace{1 mm}
\noindent but, from {\bf A.} 2, the constraint $\partial S/\partial u=0$ implies
\begin{eqnarray*}
A=(1-u)^{1/n}\frac{2\pi n}{n-1}(\frac{u}{B})^{1/(n-3)} \mbox{.}
\end{eqnarray*}
\hspace{12 cm } {\bf A.} 4

\vspace{1 mm}
Now, using $u=B r_n^{n-3}$, we get, from {\bf A.} 3, and {\bf A.} 4,

\begin{eqnarray*}
T_{0}=(2 \pi r_n)^{-1} \mbox{,}
\end{eqnarray*}
\hspace{12 cm } {\bf A.} 5

\vspace{2 mm}
\noindent which coincides with the temperature of the conpact instanton.

\vspace{1 cm}
\noindent {\bf Apendix B}

\vspace{3 mm}
If we take into account Eq. (30), i.e., $1/n=(a-1)/(d+2a-3)$,
Eq. (28) becomes,
\begin{eqnarray*}
\omega'=\frac{u'^{(d-1)/(d-a)}}{1-(1-u')^{(d-1)/d}} \mbox{,}
\end{eqnarray*}
\hspace{12 cm } {\bf B.} 1

\noindent if we define $\gamma\equiv [1-x/d]^d$, then $u'\approx \exp[-\gamma]$ and 
{\bf B.} 1  may be conveniently approximated by (assuming $\gamma\ll 1$),

\begin{eqnarray*}
\omega'\approx \exp [-\gamma\frac{d-1}{d-a}+\gamma^{(d-1)/d}] \mbox{,}
\end{eqnarray*}
\hspace{12 cm } {\bf B.} 2

\noindent that is, since semiclassically $a\sim x$, neglecting terms which are 
$O[\gamma^2]=O[(u'-1)^2]$, we get
\begin{eqnarray*}
\omega'\sim \exp [\frac{1}{d}[1-x/d]^{d-1}]+ O[(u'-1)^2]   \mbox{.}
\end{eqnarray*}
\hspace{12 cm } {\bf B.} 3

On the other hand, in order to obtain Eq. (42) 
we use, as a first approximation, the solutions of Eq. (32)
for $d=2$, i.e., $x(a,d)\sim x(a,2)=4(a-1)/a$, then
\begin{eqnarray*}
u'\approx\exp\{-[1-\frac{4(a-1)}{d a}]^d\}  \mbox{,}
\end{eqnarray*}
\hspace{12 cm } {\bf B.} 4

\noindent at $d=4$ it provides $u'\approx\exp[-1/a^4 ]$ which
is correct up to  an approximation of $1\%$. Now, we expand Eq. {\bf B.} 4 in   
a series of $(1/d)^k$ near $d=4$, to obtain

\begin{eqnarray*}
u'\sim \exp\{-\frac{1}{a^3}[1-\frac{x(a,d)}{d}+ O[(1/d)^2]] \} \mbox{,}
\end{eqnarray*}
\hspace{12 cm } {\bf B.} 5

\noindent here,  we have  replaced $x(a,2)$ by $x(a,d)$.  Eq. (42) is
obtained if we  further approximate $x(a,d)\sim a$.
Since, on the other hand, $u'\approx \exp[-\gamma]$, Eq. {\bf B.} 5  
allows us  to write the following useful expression:

\begin{eqnarray*}
[1-\frac{x(a,d)}{d}]^{d-1} \sim \frac{1}{a^3} +O[(1/d)^2]\mbox{.}
\end{eqnarray*}
\hspace{12 cm } {\bf B.} 6

\noindent Finally, Eqs. {\bf B.} 3 and {\bf B.} 6, lead to Eq. (41)


\vspace{1 cm}
\begin{thebibliography}{99}

\bibitem{kn:Gibbons1}
G.W. Gibbons and, J.B. Hartle, {\em Phys. Rev.} D{\bf  42}, 2458 (1990).


\bibitem{kn:Eguchi}
T. Eguchi, P. B. Gilkey and, A. J. Hanson, {\em Phys. Rep.} {\bf 66}, 213 (1980).

\bibitem{kn:Hawking1}
S.W. Hawking, {\em Nucl. Phys.} B{\bf  239}, 257 (1984).

\bibitem{kn:Vilenkin1}
A. Vilenkin, {\em Phys. Rev.} D{\bf 27}, 2848 (1983).

\bibitem{kn:Hartle}
J.B. Hartle and, S.W. Hawking {\em Phys. Rev.} D{\bf 28}, 2960 (1983).

\bibitem{kn:Hawking2}
S.W. Hawking and R. Penrose, {\em The Nature of Space and Time},
Princeton University Press, (Princeton, NJ) (1996).

\bibitem{kn:Hawking3}
S.W. Hawking, {\em Nucl. Phys.} B{\bf  144}, 349 (1978).

\bibitem{kn:Zeh}
H.D. Zeh, {\em The Physical Basis of the Direction of Time},
Springer, (Berlin) (1992)

\bibitem{kn:Jeans}
J. Jeans, {\em Philos. Trans. Rev. Soc. London} A{\bf  199}, 491 (1902).

\bibitem{kn:Gibbons2}
G.W. Gibbons and, M. J. Perry, {\em Proc. R. Soc. London} A{\bf  358}, 467 
(1978).

\bibitem{kn:Embacher}
F. Embacher, {\em Class. Quantum Grav.} {\bf 13}, 921 (1996).

\bibitem{kn:Gasperini}
M. Gasperini, {\em Phys. Lett.} B{\bf 224}, 49 (1989); M. Gasperini and,
G. Ummarino, {\em Phys. Lett.} B{\bf  266}, 275 (1991).

\bibitem{kn:Bekenstein}
J.D. Bekenstein, {\em Phys. Rev.} D{\bf 49}, 1912 (1994).

\bibitem{kn:York}
J.W. York, {\em Phys. Rev.} D{\bf 33}, 2092 (1986).

\bibitem{kn:Perry}
D.J. Gross, M.J. Perry, and L.G. Yaffe, {\em Phys. Rev.}, D{\bf 25}, 
330 (1982).

\bibitem{kn:KieferZeh}
C. Kiefer and  H.D. Zeh, {\em Phys. Rev.} D{\bf 51}, 4145 (1995).

\bibitem{kn:Penrose}
R. Penrose, in {\em Quantum Gravity 2}, edited by C. J. Isham,
R. Penrose, and D.W Sciama (Clarendon Press, London 1981).


\end{thebibliography}
\end{document}